\documentstyle[11pt,epsf,psfig]{article}
\def\vav{\bar v}
\def\Fc{F_{\rm c}}
\def\geta{g~\eta_{i,z_i}}
\def\zf{z^{*}}

\title{A solvable model of interface depinning 
in~random~media}
\author{Jean Vannimenus   \ and \ Bernard Derrida
   \\[1mm] 
Laboratoire de Physique Statistique de l'ENS,\\
       24 rue Lhomond, 75005 Paris, France\thanks
	 { Laboratoire associ\'e au CNRS et aux Universit\'es Paris~VI et
	 Paris~VII.
	 }}
\date{Version of \today}

\begin{document}

\maketitle

\begin{abstract}
We study the mean-field version of a model proposed by  Leschhorn to describe
 the depinning transition  of interfaces in random media.  We show that 
evolution equations for the distribution of  forces felt by
the interface sites can be written directly  for  an infinite system. 
 For a flat distribution of random local forces the  value of the
depinning threshold can be obtained exactly.
 In the case of parallel dynamics (all unstable sites move
simultaneously), 
due to the discrete character of the interface 
heights allowed in the model,
the motion of the center of mass is non-uniform 
in time in the moving phase close to the threshold, 
and the mean interface velocity  vanishes  with a square-root singularity.

\medbreak

\noindent PACS numbers: 64.60.Lx, 05.40.+j, 05.70.Ln

\medbreak

\noindent Keywords: nonequilibrium phase transition; interface; depinning.

\medbreak

\end{abstract}
\section{Introduction}

The problem of how a deformable object moves through a heterogeneous 
medium arises in many different contexts, as exemplified 
by the title of a recent review of the subject:
"Collective transport in random media: from superconductors to
earthquakes" \cite{Fisher.Collective.98}.
The displacement of a domain wall in a disordered ferromagnet
\cite{Bruinsma.Interface.84},  of an interface between two fluids in a
porous medium  \cite {Koplik.Porous.85, Levine.Porous.91} or of the contact
line of a fluid partially wetting a heterogeneous  substrate 
\cite{Rolley.Contact.98},   may  be viewed as examples of interfaces with different
elastic properties  submitted to the competing effects of  an external driving force
and of local random pinning forces
\cite{Fisher.Collective.98, Nattermann.Renorm.92, Nattermann.pinning.97,
Kardar.dynamics.98}.

 A general feature of these  systems is the existence of a
depinning threshold: below a well-defined external force $\Fc$ 
the interface  does not move (at zero temperature), due to the collective
action of the pinning centers, while for $F > F_{\rm c} $ it moves with 
a mean  velocity~$\vav$. 
Close to the threshold this velocity has a singular behaviour, 
analogous to a critical phenomenon:
\begin{equation}
 \vav \sim  (F- F_{\rm c})^{\theta} .
\label{eqn:theta}
\end{equation}
Field theory methods such as the dynamic functional renormalization group 
\cite{Nattermann.Renorm.92, Narayan.Interfaces.93, Balents.Expansion.93,
Kardar-Ertasz.lines.94, Chauve.Creep.2000} have been used to predict the dependence of
$ \theta$   and other critical exponents  on the space dimensionality, 
the nature of the disorder  and the range of  the  interactions 
between parts of the interface.  
The replica method has also been applied to the problem, indicating that
typical pinned interfaces have the same roughness properties as slowly moving ones
\cite{Mezard.Replicas.99}. 
 
 Exact results, even for simple models of the depinning transition, have not been
obtained so far, 
 and our goal is to show how a class of mean-field models can be solved exactly.

 We consider 
a model first introduced by Leschhorn
 \cite{Leschhorn.Model.92}, 
where the disorder is of the random force type and where space and time are
 discrete,  which appears to give a good qualitative description of that
transition
\cite{Leschhorn.Numerical.93, Jost.Numerical.98}. 
We first show that 
for the mean-field version of that model
dynamical equations can be written  exactly in the thermodynamic limit  
for parallel (i.e., synchronous) dynamics. 
From these equations of evolution  the threshold can be
obtained explicitly for some distributions of the random local forces. As an
example we study  the specific case of a flat distribution and obtain the
corresponding value of the threshold as the solution of an algebraic equation. 
Numerical results are presented for the distance travelled by the interface before
stopping, in the pinned phase close to threshold.

 We next consider the motion of the interface in the moving phase
and find that the  
  mean velocity 
vanishes as a power law (\ref{eqn:theta}), with an exponent  $\theta = 1/2$.
 This value of $\theta$ differs from the values found 
for models with continuous space and continuous relaxational dynamics,
for which the mean-field behaviour  depends on the 
form  of the pinning potential
 \cite{Narayan.CDW.92, Narayan.Interfaces.93}.
We discuss the origin of this difference and  relate it
to the non-uniformity of the motion very close to the threshold, 
 which  is usually not taken into account 
in mean-field theories. 
This non-uniform motion is itself an effect of the discreteness of
the allowed heights for the interface.

\section{Model and evolution equations}

\subsection{The Leschhorn models}

 Different models have been proposed to describe depinning phenomena, with
varying degrees of realism, but in view of the very general nature of the
problem, it is instructive to study its main qualitative features on the
simplest possible systems.
In this spirit Leschhorn 
\cite{Leschhorn.Model.92, Leschhorn.Numerical.93}
introduced models discrete in both space and time,
 in which the interface evolves according to very simple rules. 
These models are  easier and faster to simulate than more realistic
 continuous models and allow a detailed study of the interface motion
 near the threshold. 

 As usual in such studies it is assumed  that 
overhangs may be neglected near the threshold, so 
a unique height $z_i(t)$  is associated with 
 each site $i$ ($1 \le i \le  N$) of the discretized   
$(d-1)-$dimensional interface.   
By definition of the model the total force $f_i$ 
on this site is the sum of the restoring force from  the other sites, 
of an external force $F_{ext}$  and of a random local force: 
\begin{equation}
f_{i}(t) = \sum_{j}K_{i,j}\Bigl[z_{j}(t) - z_{i}(t)\Bigr] + F_{ext} + \geta \; ,
\label{eqn:Lesch1}
\end{equation}
 and at each time step the interface may move forward 
 at site $i$ only if  $f_i$  is positive, otherwise it does not move.

The interactions $ K_{i,j} $ in (\ref{eqn:Lesch1}) are positive (or zero), so a
site which lags behind its neighbours experiences a positive restoring force  and it
will move unless the pinning force on it is sufficiently negative.
The coupling constant $ g $  fixes the scale of the random force,  
while the  $\eta_{i,z_i}$ are uncorrelated random numbers drawn
 from a given normalized distribution $\rho(\eta)$, 
a new random number being  drawn only if the interface moves at site $i$
 (this is what distinguishes pinning models from random growth models
described by similar equations, but where 
a new random number is drawn at each time step).  

 As will be shown in the following, the time evolution is described by
relatively simple equations in the case  where all sites
experiencing a positive force are updated simultaneously (parallel
dynamics) and  move by  one lattice unit, independently of the magnitude of that
force.  The equations of motion then read

\begin{equation}
z_{i}(t+1) = \left\{ \begin{array}{ll} 
                z_{i}(t) + 1  &  \mbox{\quad  if  \quad $f{_i}(t) > 0 , $ } \\
                z_{i}(t)      &  \mbox{\quad  if  \quad $f{_i}(t) \le 0 .$ }
                  \end{array} 
            \right.
\label{eqn:Lesch2}
\end{equation}
As a consequence the instantaneous velocity of the center of mass is just equal to the
fraction of sites with a strictly positive force $f{_i}$. Another
interesting case  considered in the literature is that of "extremal dynamics"
\cite{Nadal.extrem.86, Roux.dynamique.94, Tang.dynamics.94}, where  at each time step
only the site with the largest positive value of $f_i$  moves. 
  
Different interaction kernels  $K_{i,j}$ correspond to various physical
situations:  
for an elastic interface   $K_{i,j} \ne 0$ only if  $i$  and  $j$  are
neighbouring sites,  while in the contact line problem the interaction decays 
slowly  with the distance  $(i-j)$ 
\cite{Pomeau.Contact.85, deGennes.Contact.85}.
The model defined in (\ref{eqn:Lesch1}) and (\ref{eqn:Lesch2}) has been
studied numerically by Leschhorn for nearest-neighbour interactions,
  in the case where $\eta$ can take two values,
$+1$ and $-1$, with probabilities $p$ and $1-p$  respectively
\cite{Leschhorn.Numerical.93, Nattermann.pinning.97}. 
For $d=2$ (a 1-d interface moving in 2-d space), the results were 
 rather noisy and a large computing effort was needed to obtain an estimate 
$\theta~=~0.25~\pm~0.03$.
For $d=3$ the value obtained,
$\theta~=~0.64~\pm~0.02$,  
is to be compared with  the prediction 
 $\theta~=~2/3$  
obtained by a first-order expansion about the upper
critical dimension, $d_{\rm c}=5$ for short-range interactions
\cite{Nattermann.Renorm.92}. 

\vspace{5mm}

{\it Remarks:}

- A non-zero mean value of the distribution $\rho(\eta)$
has the same effect as an external force $g <\eta>$,
so one can assume that $F_{ext}=0$ without loss of generality.
Note that, due to the asymmetry between positive and negative forces in
(\ref{eqn:Lesch2}), a transition exists even in the absence 
of an external force
  and for $\rho(\eta)$  symmetric   (i.e., $\rho(\eta) = \rho(-\eta)$ ). 
The calculations below are carried out with $F_{ext}=0$, 
for notational simplicity,  except when indicated.

- The "no-passing" theorem holds \cite{Middleton.passing.93}: if  interface $A$ is
everywhere ahead of interface $B$, i.e., 
$ z_i^{A}(t) \ge  z_i^{B}(t) $  for all $ i $, 
then at all their contact points 
$ f_i^{A}(t) \ge  f_i^{B}(t) $,
so $B$ cannot pass $A$ on these sites. At the remaining  points $B$ can at best
catch up with $A$, since it moves only one lattice unit at a time. 
As a consequence moving and static interfaces cannot
coexist in the same sample. 
 
-  In the thermodynamic limit ($N \rightarrow \infty$), 
the interface moves forward indefinitely for small  values of 
$g$, and it is pinned for large values of $ g $.
 In that limit  there exists a critical coupling $g{_c}$,   
 such that for $g < g_c$ the interface moves with a non-zero mean velocity. 
The threshold $g{_c}$ plays the same role as  $F{_c}$ in (\ref{eqn:theta}), 
and we are interested in what happens for
 $g$  close to $g{_c}$.

\subsection{Mean-field evolution equations}

 The mean-field theory (MFT in the following) is usually identified with the
infinite-range limit,  where each interface site interacts equally 
with all the others \cite{Fisher.Density.85, Koplik.Porous.85}, 
 i.e.,  the total force on site $i$ is of the form
\begin{equation}
f_i(t) = \bar z(t) - z_i(t) + \geta ,
\label{eqn:MeanField}
\end{equation}
where
\begin{equation}
 \bar {z}(t) = {1\over N}\sum_{i=1}^N z_i(t)
\end{equation} 
is the average instantaneous position of the interface.

This limit has  also been studied numerically
\cite{Leschhorn.Model.92},  
for the case where $\eta$ can take three  values ($1,0, -1$).  
 It was found  that the mean interface velocity
 vanishes  linearly at the depinning threshold, i.e., $\theta= 1$,
 in agreement with 
the mean-field prediction for models with discontinuous random forces
\cite{Nattermann.Renorm.92, Narayan.CDW.92}.
  
 We will show now that in this mean-field limit one can write exact evolution
equations for $P_{k}(x, t)$, the fraction of sites  at height $k$ and experiencing a
local pinning force $ g~x$:
 \begin{equation}
 P_{k}(x,t)=  \lim_{N \rightarrow \infty} \left\{ {1\over N} \sum_{i=1}^{N}
             \delta_{z_i(t),k} \;  \delta(x -\eta_{i,z_i}) \right\} .
 \label{eqn:Proba}
 \end{equation}
It obviously satisfies the normalization condition
\begin{equation}
  \sum_k {\int_{-\infty}^{\infty} P_k(x,t) dx} = 1. 
 \label{eqn:norm} 
 \end{equation}
Note that we are considering directly  quantities defined for the infinite
system, thus avoiding the difficulties associated with finite-size effects.

 Let us consider the time $t$ when the interface first 
reaches a given height $k$.
  For the newly occupied sites at $k$,  $P_{k}(x, t)$
is just  proportional to $\rho(x)$.
At the next time step, for the parallel dynamics studied here, 
all the sites  with $f_i > 0$  
move one step forward,  so  $P_{k}(x, t)$  
gets truncated  at
 $x^{\ast} = (k - \bar {z}(t))/g$      (Figure~1). 
At the same time, among the interface sites located at height $(k -1)$, 
all those  for which $f_i > 0$  move forward,  
and new random numbers are drawn for those sites, adding to $P_{k}(x)$
 a contribution  also proportional to $\rho(x)$, for all $x$. 
This construction may be repeated for the following time steps, showing
that, for all heights  $k$ ahead of the region initially occupied by the
interface,   $P_k(x,t)$ consists of  two parts,  each of them proportional to
$\rho(x)$.

\begin{figure}
 \centerline{\psfig{figure=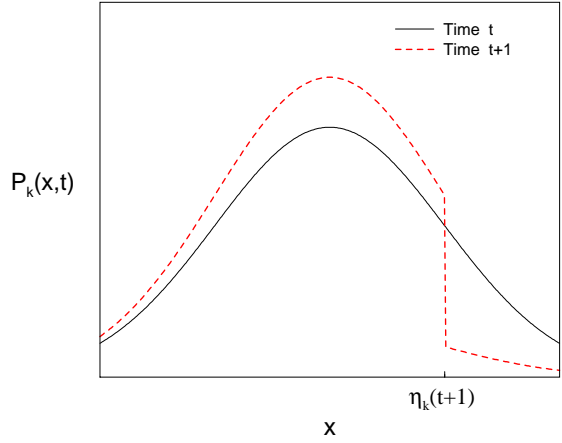,height= 6cm}}
\caption{ Probability distribution of the forces on the interface sites at height $k$,
at time $t$ (when $k$ is first occupied) and at time $t+1$.}
\label{fig_gauss}
\end{figure}

 More generally the evolution equations      
are
 \begin{equation}
 P_k(x,t+1)= \left\{ \begin{array}{ll}
       P_k(x,t) + \rho(x) {\int_{\eta_{k-1}(t+1)}^{\infty}\ P_{k-1}(x',t) dx'}
       & \mbox{\quad if  $\; x < \eta_k(t+1)$,}\\
         \\
        \rho(x)  {\int_{\eta_{k-1}(t+1)}^{\infty}\ P_{k-1}(x',t) dx'}
       &  \mbox{\quad if   $\; x > \eta_k(t+1)$,} 
            \end{array}
          \right.
 \label{eqn:distrievol}
 \end{equation}
where the discontinuity point  varies linearly with $k$, according to
\begin{equation} 
\eta_{k}(t+1)= \frac{k - \bar {z}(t)}{g}
\label{eqn:cutoff}
\end{equation}
%
(to avoid the case 
$x = \eta_{k}(t+1)$ 
we assume that $\rho(\eta)$ does not contain delta peaks). 

If initially 
 $ P_k(x,t=0)$ has the form
 \begin{equation}
 P_k(x,t=0)=  c_{k} \rho(x) , 
 \label{eqn:initial}
 \end{equation}
with $\sum_{k} c_{k} = 1$ 
(in particular if initially the interface is flat: $c_k = \delta_{k,0}$), 
then from (\ref{eqn:distrievol})  $P_k(x,t)$ may be written as
 \begin{equation}
 P_k(x,t)= \left\{ \begin{array}{ll}
           \lambda_{k}(t)~\rho(x) & \mbox{\quad if \quad $x < \eta_k(t)$,}\\
            \mu_{k}(t)~\rho(x)   &  \mbox{\quad if \quad $x > \eta_k(t)$.} 
            \end{array}
          \right.
 \label{eqn:distrib}
 \end{equation}
Due to this  simple structure 
the evolution equations (\ref{eqn:distrievol}) can be expressed in terms of the
$\lambda_{k}(t)$ and $\mu_{k}(t)$
\begin{eqnarray}
\mu_{k}(t+1) & = & \lambda_{k-1}(t) {\int_{\eta_{k-1}(t+1)}^{\eta_{k-1}(t)}
\rho(x) dx} + \mu_{k-1}(t) {\int_{\eta_{k-1}(t)}^{\infty}\rho(x) dx},  
\label{eqn:muk2}  \\
\lambda_{k}(t+1) & = & \lambda_{k}(t) +  \mu_{k}(t+1).
\label{eqn:lambdak2}
\end{eqnarray}
The initial conditions are
 \begin{equation}
 \lambda_k(0)=  \mu_{k}(0) = c_{k}, 
 \label{eqn:initialbis}
 \end{equation}
and from  (\ref{eqn:lambdak2}) 
$\lambda_{k}(t)$ is a non-decreasing function of time.

 Both $\lambda_k(t)$ and  $\mu_{k}(t)$ have a simple physical interpretation. 
From (\ref{eqn:muk2}), $\mu_{k}(t+1)$ is the fraction of sites which jump from height
$k-1$ to height $k$ at time $t+1$. Then  it is clear from (\ref{eqn:lambdak2}) that
 \begin{equation}
 \lambda_{k}(t) = \lambda_{k}(0) + \sum_{t'=1}^{t} \mu_{k}(t'), 
 \label{eqn:lambdatot}
 \end{equation}
so $\lambda_{k}(t)$ is just the total fraction of sites 
with a height $\ge k$, if initially all heights larger than $k$ are empty 
($c_{k'} = 0$ for $k' > k$).

 \subsection{ A recurrence relation on the $\lambda_k$}

  The evolution equations can be written in a form involving only
$\lambda_{k}$, by calculating $\mu_{k}$  from (\ref{eqn:lambdak2}) 
and using that expression in (\ref{eqn:muk2}). 
One gets
 \begin{equation}
\lambda_{k}(t+1) - \lambda_{k-1}(t) {\int_{\eta_{k-1}(t+1)}^{\infty} \rho(x) dx}
 = \lambda_{k}(t) - \lambda_{k-1}(t-1) {\int_{\eta_{k-1}(t)}^{\infty} \rho(x) dx}.
\end{equation}
 The two sides of this equality correspond to the same quantity at two
successive times, so it is independent of time and equal to its value at $t = 0$, 
$\lambda_{k}(0) (= c_k)$.
  One obtains 
\begin{equation}
 \lambda_{k}(t+1) =  c_{k} + \lambda_{k-1}(t) {\int_{\eta_{k-1}(t+1)}^{\infty}
 \rho(x)} dx ,
 \label{eqn:lambdat}
\end{equation}
where we recall (see (\ref{eqn:cutoff})) that $ \eta_{k} $  satisfies
\begin{equation} 
\eta_{k}(t+1)= \frac{k - \bar {z}(t)}{g}.
\label{eqn:cutoffbis}
\end{equation}

 For an interface initially flat and located at $k = k_0$, 
 $\lambda_{k}$ is the fraction of sites at heights
$\ge k$, so the average interface height is just given by
\begin{eqnarray}
\bar{z}(t) & = & \sum_{k \ge k_0} {k \: \left[ \lambda_{k}(t) - \lambda_{k+1}(t)
\right]}, 
\label{eqn:ztlambda}
\\ 
  & = &  k_0 + \sum_{k > k_0} {\lambda_{k}(t)}.
 \label{eqn:zt1}
\end{eqnarray}
For the more general initial conditions (\ref{eqn:initialbis}) this relation becomes
\begin{equation} 
\bar{z}(t) - \bar{z}(0) = \sum_{k} \left[ {\lambda_{k}(t) - c_k} \right].
\label{eqn:ztgen}
\end{equation}
The instantaneous velocity of the center of mass is thus given by 
\begin {equation}
v(t) = \bar {z}(t+1) - \bar {z}(t)  = \sum_{k} \left[ \lambda_{k}(t+1) -
\lambda_{k}(t)  \right] ,
\label{eqn:vt1}
\end{equation} 
a form useful for numerical purposes when this velocity is very small.

Taking the $\lambda_{k}(t)$ as basic variables, 
$\bar {z}(t)$ and $\eta_{k}(t+1)$ can be obtained from 
 (\ref{eqn:cutoffbis}) and (\ref{eqn:zt1}) or (\ref{eqn:ztgen}).
So, together with (\ref{eqn:lambdat}) and the initial conditions
(\ref{eqn:initialbis}),
these equations constitute a  dynamical system  describing the
average evolution of an interface, in the thermodynamic  limit. 
This system can be used as such for numerical studies, 
with the advantage over  conventional Monte Carlo simulations that
finite-size effects as well as numerical noise are absent. 
This makes it possible
to determine the threshold and to study the critical
properties with a much better  accuracy.

\vspace{5mm}

{\it Remarks:}

- For a deformable object like an interface the notion of
instantaneous velocity is not unique: 
For example one may consider the velocity of the center of mass, or 
of the leading edge, and they are usually different.  
In the present model the velocity of the leading edge fluctuates strongly, being
either $0$ or $1$,  but close to the threshold the time interval between two
non-zero values increases.
 Of course, its time average is identical to the time average of the 
 velocity of the center of mass and it vanishes at the threshold.

- Exact dynamical mean-field equations for an infinite system have been
obtained  by Eissfeller and Opper \cite{Opper.Infinite.92} 
for spin glasses, but in their approach the resulting equations still contain a
noise term and have to be solved by Monte Carlo simulations.

\section{An analytically  solvable case}

For some distributions $\rho(\eta)$ of the local random forces,
the integrals that appear in (\ref{eqn:lambdat}) can be  calculated easily, 
making it possible to push the analytical study further.
As a simple example 
we treat the  case of a flat symmetric distribution:
\begin{equation}
\rho(\eta) = \left\{ \begin{array}{lc}
                   1/2 & \mbox{\quad if $-1 \le \eta \le 1$,} \\
                    0  & \mbox{\quad otherwise}
                  \end{array}
             \right. 
 \label{eqn:rhoflat}
\end{equation}
(any symmetric flat distribution may be reduced to (\ref{eqn:rhoflat})  by
rescaling the coupling constant $g$). For simplicity we consider in the following
only a flat interface initially located at $k=0$.

\subsection{Evolution equations}

   $P_k(x,t)$ is then just made up of two constant parts
 and is non-zero  only for a finite number of  values of $k$,
as only a finite number of the $\eta_{k}(t)$  
given by (\ref{eqn:cutoffbis})
 lie in the interval $(-1, 1)$ where $\rho(\eta)$ is non-zero. 

Consider first the leading occupied edge: it stays at $k =  k_{max}(t)$ 
if the total force (\ref{eqn:MeanField}) on each of its sites is $\le 0$,
i.e., if $ k_{max}(t) - \bar {z}(t) \ge  \ g $. 
If not, it moves to $ k_{max}(t) + 1$, so one has the bounds
\begin{equation}
    \bar {z}(t) + g +1 \; > \;  k_{max}(t+1) \; \ge \; \bar {z}(t) + g.
 \label{eqn:kmax} 
\end{equation}
 As for the trailing edge, its position  $k_{min}(t)$ remains fixed 
as long as its most strongly pinned sites  
experience a non-positive total force, i.e., if 
$\bar {z}(t) \le k_{min}(t) + g$.
It moves to $k = k_{min}(t) +1$ if on all its sites $f_i > 0$. 
These two conditions imply that
\begin {equation}
  \bar {z}(t) - g +1 \; > \;  k_{min}(t+1) \; \ge \; \bar {z}(t) - g,
\label{eqn:kmin}
\end{equation} 
and combining (\ref{eqn:kmax}) and (\ref{eqn:kmin}) one obtains bounds 
on the interface width:
\begin {equation}
   2 g + 1  \ge  \;  k_{max} -  k_{min} \;  \ge 2 g - 1 .
\label{eqn:width}
\end{equation} 

The evolution equations 
(\ref{eqn:lambdat})  become
\begin {equation}
 \lambda_{k+1}(t+1)  = 
   \left\{ \begin{array}{lc} 
          \lambda_{k}(t) \; \frac{1-\eta_{k}(t+1)}{2} & 
                 \mbox{if}  \quad   \vert \eta_{k}(t+1) \vert \le  1, \\
        \\
           0  & \quad \mbox{if}  \quad \eta_{k}(t+1) \ge 1, \\
        \\
          \lambda_{k}(t)  & \quad \mbox{if} \quad \eta_{k}(t+1) \le -1,
        \end{array}
   \right.     
 \label{eqn:lambdat2}   
\end{equation} 
with $\lambda_{0}(t) \equiv 1$ and $\bar {z}(t=0) = 0$ as boundary conditions. 
 The position of the leading  edge
 is such that $\lambda_{k}(t) = 0$ for $k > k_{max}(t)$,
  and the position  $k_{min}(t)$ of the trailing edge is
  the largest value of $k$ such
  that  $\lambda_{k}(t) = 1$
  (remember that $\lambda_{k}$ is the fraction of sites at heights
  larger \emph{or equal} to $k$). 
 The average interface position, eq.(\ref{eqn:zt1}), may then be expressed
as
\begin {equation}
  \bar {z}(t) = \; k_{min}(t) \; + \sum_{k>k_{min}(t)} \lambda_{k}(t),
  \label{eqn:zlambda}
\end{equation}
a form that will be useful in the following.

\subsection{Determination of the depinning threshold}

In order to find the threshold $g_c$ we note that very close to it,
in the moving phase, the interface velocity is very small (if the transition is
continuous, as found numerically), implying from (\ref{eqn:vt1}) that
$\lambda_{k}(t+1) \simeq \lambda_{k}(t)$  for all $k$. 
One is therefore close to a fixed point of 
(\ref{eqn:cutoffbis}), (\ref{eqn:zt1}) and (\ref{eqn:lambdat2}).
But if a 
fixed point exists for a given value of the coupling $g$, 
there is a non-moving solution of the evolution equations and according to the
no-passing theorem  this is incompatible with the assumption that the system 
is in the moving phase. 
This shows that the threshold $g_c$ corresponds to the appearance of a fixed
point when $g$ is increased: For $ g < g_c$, there is no fixed point and the
interface moves indefinitely; for $ g > g_c$, there is a fixed point and the
interface comes to a halt. 

\subsubsection{Self-consistent equations for the halted interfaces}

 The halted solutions of the evolution equations satisfy 
the following self-consistent system:
\begin{eqnarray}
 \eta_{k} & = & \frac{k - \zf}{g}, 
  \label{eqn:etafix} \\ 
 \lambda_{k+1} & = & \lambda_{k} \; \frac{1-\eta_{k}}{2}
   \qquad \mbox{for} \quad \vert \eta_{k} \vert \le  1, 
 \label{eqn:lambdafix}\\
 \zf & = &  k_{min} + \sum_{k>k_{min}} {\lambda_{k}},
  \label{eqn:pfix}
\end{eqnarray}
with 
 \begin {equation}
   \lambda_{k} = 1 \quad \mbox{for} \quad  k \leq k_{min}, 
   \quad \; \lambda_{k} = 0 \quad \mbox{for} \quad  k \geq \zf + g +1 .
 \end{equation}
In these equations $\zf$ and $k_{min}$ denote respectively the
positions of the center of mass and of the trailing edge of a  halted
interface, whose distribution of local forces (\ref{eqn:distrib}) is given by
the $\lambda_{k}$ and $\eta_{k}$,  with $\mu_{k} =0$. 

 Noting that (\ref{eqn:lambdafix}) may be written
\begin {equation}
  \lambda_{k} - \lambda_{k+1} =  \lambda_{k} \frac{1+\eta_{k}}{2} ,
 \end{equation}
and using (\ref{eqn:ztlambda}) to reexpress $\zf$, 
one checks easily that a solution of the above
system verifies both the normalization and the self-consistency conditions,
 which here are simply
\begin {eqnarray}
 \sum_k {\int_{-\infty}^{\infty} P_k(x,t) dx} & = & \sum_{k \ge k_{min}} {\lambda_{k}
\frac{1+\eta_{k}}{2} } \quad = \quad 1, \\
  \sum_k {\int_{-\infty}^{\infty} k P_k(x,t) dx} & = & \sum_{k \ge k_{min}}
{k \lambda_{k} \frac{1+\eta_{k}}{2} } \quad  =  \quad  \zf.
\end{eqnarray}
Other stationary solutions verifying these two conditions 
but not (\ref{eqn:lambdafix}) 
 exist, but they would correspond to other initial conditions 
and cannot be reached starting from  a flat interface.

\subsubsection{The threshold}

 In order to find the halted solutions explicitly, we first note 
that they are invariant by
 a translation through an integer number of lattice units,  
so we can fix $k_{min} = 0 \;$ for simplicity. 
In addition, the system (\ref{eqn:etafix} - \ref{eqn:pfix})
can be reduced to an equation for the single variable $\zf$,
 as the $\eta_{k}$ may be obtained from
$\zf$ using (\ref{eqn:etafix}), then the $\lambda_{k}$ from 
(\ref{eqn:lambdafix}) and $\lambda_{0} = 1$.
Reinjecting these values into (\ref{eqn:pfix})
yields a self-consistent equation for $\zf$. 
The precise form of this equation depends on the width of the halted
interface, for which only the bounds (\ref{eqn:width}) are known. A search
for solutions can be made for the different possible values of the width, but the
effort can be reduced using hints obtained from numerical studies.
 These indicate that  the threshold is very close
 to $g=2.38$, and that the width of the halted interface is 
 $(k_{max} - k_{min}) = 5$.
%
%

Introducing for convenience the variables 
\begin {equation}
y = (1- \eta_{0})/2 \quad \mbox{and} \quad  u = 1/{2g},
\end{equation}
equation (\ref{eqn:etafix}) may be written
\begin{equation} 
(1 - \eta_{k})/2 = y - k \; u ,
\label{eqn:vary}
\end{equation}
and from (\ref{eqn:lambdafix}) and (\ref{eqn:pfix})
the self-consistency conditions  for a halted interface of
width $5$  read
\begin {equation}
\zf  = g \; (2 y -1) = \sum_{k=1}^{5} \lambda_k ,
\label{eqn:consistency}
\end{equation}
where the $\lambda_k$, considered as functions of $y$ and $u$, 
are  polynomials of degree $ k $ in $ y $,  defined by
\begin {eqnarray}
\lambda_1 = y , \qquad \lambda_{k+1} = (y - k \; u) \lambda_k .
\end{eqnarray}
Eq. (\ref{eqn:consistency})  finally reduces to a polynomial equation 
of degree 5 in $y$ :
\begin {eqnarray}
R_{g}(y) &=& g + y [1-2g-u+2u^2-6u^3+24u^4] + y^2 [1-3u+11u^2-50u^3]
\nonumber \\
&  &  + y^3 [1-6u+35u^2] + y^4 [1-10u] + y^5  \quad = \quad  0 .
\label{eqn:polynome}
\end{eqnarray}
All the $\lambda_k $ 
have to be positive, so
acceptable solutions lie in the range $2/g<y<1$.
 They  exist only if $ g \ge g_c $ , 
the value of $ \ g $ for which $R_{g}(y)$ admits a double root in that range,
  which we identify with the depinning threshold.

Determining if a polynomial has a double root is  a
standard problem  in algebra \cite{Korn.algebra.61} and it may be done very
accurately.
One obtains
 \begin{equation}
g_c  =  2.38006232... ,
 \label{eqn:gcrit}
 \end{equation}
%
The other  parameters of the  critical halted interface are  
 \begin{equation}
 \eta_{0}= -0.7990787..., \quad  \lambda_1 =  y_c = 0.89953936..., \quad  
  \zf = 1.901857...  
\label{eqn:parametres}
 \end{equation}
Its density profile is given by the differences  
$ (\lambda_{k} - \lambda_{k+1})$,  which may be deduced 
from these values using  (\ref{eqn:etafix}) and (\ref{eqn:lambdafix}). 

\vspace{5 mm}

{\it Remark:}
 The value of the threshold does not depend 
on the particular dynamical rules chosen, 
as long as only unit jumps are allowed and the stopping rule
is $f{_i} \le 0$. 
 To see this, let us show that a weaker form of the no-passing theorem holds for
interfaces with different dynamics. Consider an interface pinned under parallel
dynamics:
 this implies that on all its sites the force $f_i \le 0 $, so it would also be
pinned under extremal dynamics. More generally an interface moving under extremal
dynamics cannot pass one under parallel dynamics.
 The converse is not true, 
but a parallel interface $B$ cannot pass a pinned extremal
one $A$,  since at all their contact points $ f_i^{B} \le  f_i^{A} \le  0$.
 This suffices to show that the nature of the phase for a given value of $g$ 
is the same  for both types of dynamics, 
 though some aspects of the critical behaviour might depend on the dynamics
considered \cite{Paczuski.SOC.96}, \cite{Narayan.anomalous.00}. 

\subsubsection{Effect of an external force}

 The case of a non-zero external force can be treated along similar lines. 
The final result is that the threshold now depends on $F_{ext}$ and
is   given by the value of $g$ for which the equation 
\begin {equation}
R_{g}(y) + F_{ext} =0,
\label{eqn:polyF}
\end{equation}
has a double root.
The expression (\ref{eqn:polynome}) of $R_{g}(y)$ may change when 
$F_{ext}$ varies, as it depends on the width of the
critical interface, which in turn may depend on $F_{ext}$.
 For small  $F_{ext}$, however, we expect the width to remain the same (= 5) 
and since at the threshold for zero external force
\begin{equation}
R_{g_c(0)}(y_c) = \partial R_{g_c(0)}(y)/ \partial y \mid _{y= y_c} =0,
\end{equation}
one obtains by expanding (\ref{eqn:polyF}) a linear dependence of 
 $g_c$ on $F_{ext} \; $  :
\begin{equation}
g_{c}(F_{ext}) - g_{c}(0) \simeq  - \frac{F_{ext}}{(\partial R_{g}/ \partial g)^{\ast}}
 \simeq  2.3901... \; F_{ext} \; ,
\end{equation}
 where the star symbol denotes a quantity taken at the threshold for  $F_{ext} = 0$. 

\subsection{Stopping distance in the pinned phase}

Armed with these exact results, we can now study in detail the behaviour 
of  (\ref{eqn:cutoffbis}), (\ref{eqn:lambdat2})   and (\ref{eqn:zlambda})
in the immediate vicinity of the threshold.

Figure~2 displays numerical results for the distance $z_{f}(g)$ at which the
center of mass of an interface initially located at $k=0$ stops, in the pinned phase
just above $g_c$. 
When $g \rightarrow g_c$,  the data are well fitted by 
\begin {equation}
  z_{f}(g)  \simeq \zf - c \sqrt{g - g_c},
\label{eqn:zfinal}
\end{equation}
where  $\zf$ is  the value (\ref{eqn:parametres}) obtained for the critical
interface (this should be expected since  at $g_c$ there exists only one stationary
solution with $k_{min} = 0$).

\begin{figure}
 \centerline{\psfig{figure=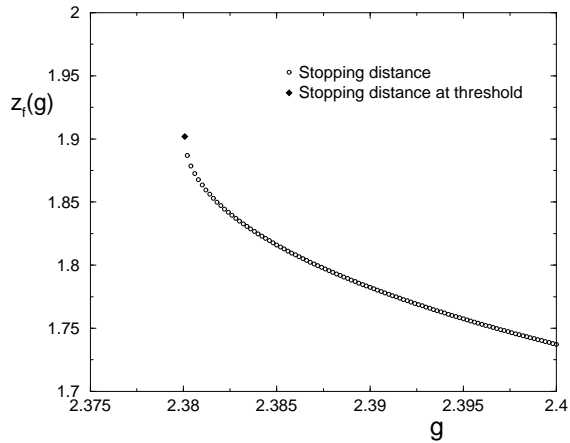,height=6cm}}
 \caption{Measured stopping distance $z_{f}(g)$ for an interface initially flat at position $z=0$,
in the pinned phase, for a uniform distribution of random local forces. The filled
diamond indicates the exact values ($z^*,~g_c $) obtained in (\ref{eqn:gcrit}) and
(\ref{eqn:parametres}).}
\label{fig_pfixg}
\end{figure}

It is also interesting to study the way the interface slows down before stopping. 
The numerical results show that its velocity vanishes linearly as a function of 
$\bar {z}(t)$ :
\begin {equation}
  v(t)  \simeq  a \;  (z_{f}(g) - \bar {z}(t) ).
\label{eqn:vslowing}
\end{equation}
When $v(t)$ is very small it may be replaced 
by its continuous approximation $d\bar{z} / dt$, and integration of 
(\ref{eqn:vslowing}) yields 
an exponential convergence at large times:
\begin {equation}
  \bar {z}(t)  \simeq z_{f}(g) - c'  \; e^{- a t}.
\label{eqn:zslowing}
\end{equation}

The factor $a$ in (\ref{eqn:vslowing}) depends on $g$, 
it vanishes when $g \rightarrow g_c$  and right at threshold convergence is found to
be much slower.  
Figure~3 shows  $[v(t)]^{1/2}$ versus $\bar {z}(t)$ at $g = g_c$. 
The data are well fitted by
\begin {equation}
  v(t)  \simeq  b\; (\zf - \bar {z}(t))^2,
\label{eqn:vcrit}
\end{equation}
from which one obtains that asymptotically 
\begin {equation}
 \bar {z}(t)  \simeq \zf - 1/ b t.
\label{eqn:zslow}
\end{equation}

\begin{figure}
 \centerline{\psfig{figure=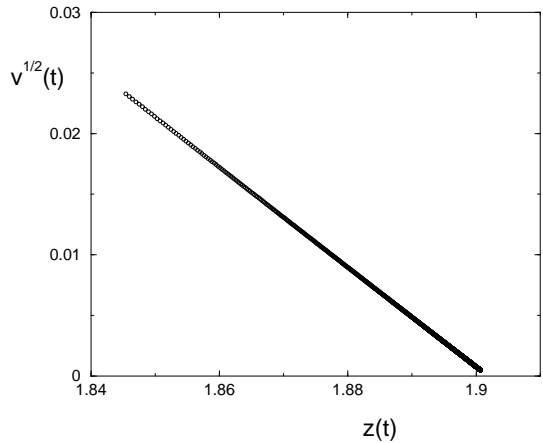,height=6cm}}
 \caption{Instantaneous velocity of the center of mass {\it vs} its position, at the
 depinning threshold ($g_c = 2.3800623...$).}
 \label{fig_gstop}
\end{figure}

As will be argued in the next section and developed in the Appendix, 
these results can be simply understood in the framework of 
a standard saddle-node bifurcation for dynamical systems.

\subsection{Moving phase: Numerical results and heuristic arguments}

\subsubsection{Numerical results}

In the moving phase, 
when the coupling $g$ increases starting from low values, 
one observes numerically  that, for $g <  g_c - 0.02$, the mean
velocity first decreases linearly  with $g$.
But this behaviour changes   
closer to the threshold (Figure~4). 
Very close to the threshold the time dependence of $v(t)$,
the instantaneous velocity of the center of mass, 
also becomes non-uniform: 
most of the time the interface  creeps very slowly, with sudden bursts 
during which it moves much more rapidly. 
The motion looks periodic  
(Figure~5), with a spatial period of one lattice unit. 
\begin{figure}
 \centerline{\psfig{figure=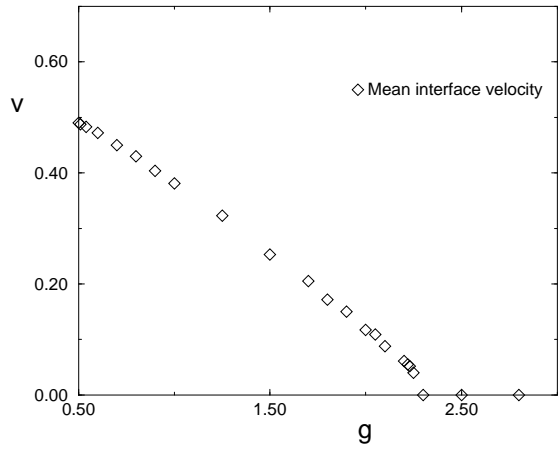,height=6cm}}
 \caption{Average interface velocity {\it vs} $g$, in the moving phase.}
 \label{fig_vgmov}
\end{figure}

\begin{figure}
 \centerline{\psfig{figure=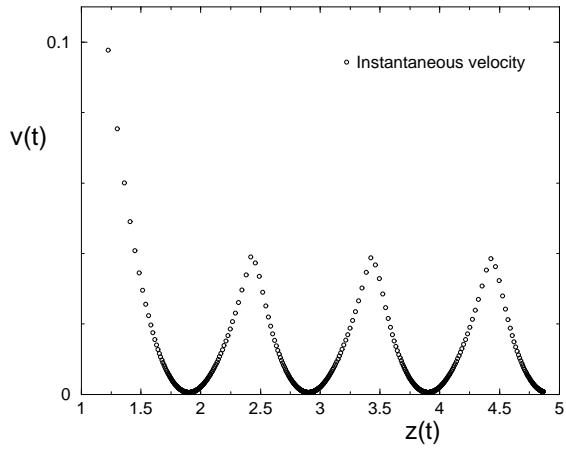,height=6cm}}
 \caption{Instantaneous interface velocity {\it vs} average position, at $g=2.378$, 
  very close to the threshold.}
 \label{fig_vz378}
\end{figure}

The \emph{minimum} velocity of the center of mass vanishes
 linearly with $(g_c - g)$ up to the threshold, 
but its \emph{mean} velocity, 
measured over a time sufficiently long to cover several lattice units, 
decreases like $(g_c - g)^{1/2}$  (Figure~6).
This  behaviour would be difficult to observe in a Monte Carlo study 
of the model defined by  (\ref{eqn:Lesch2}) and (\ref{eqn:MeanField}), 
due to finite size effects
and to the increasing period. 
  In particular, stopping the simulations after a fixed number of time steps,
independently of the distance to $g_c$, would give
incorrect results.

\begin{figure}
 \centerline{\psfig{figure=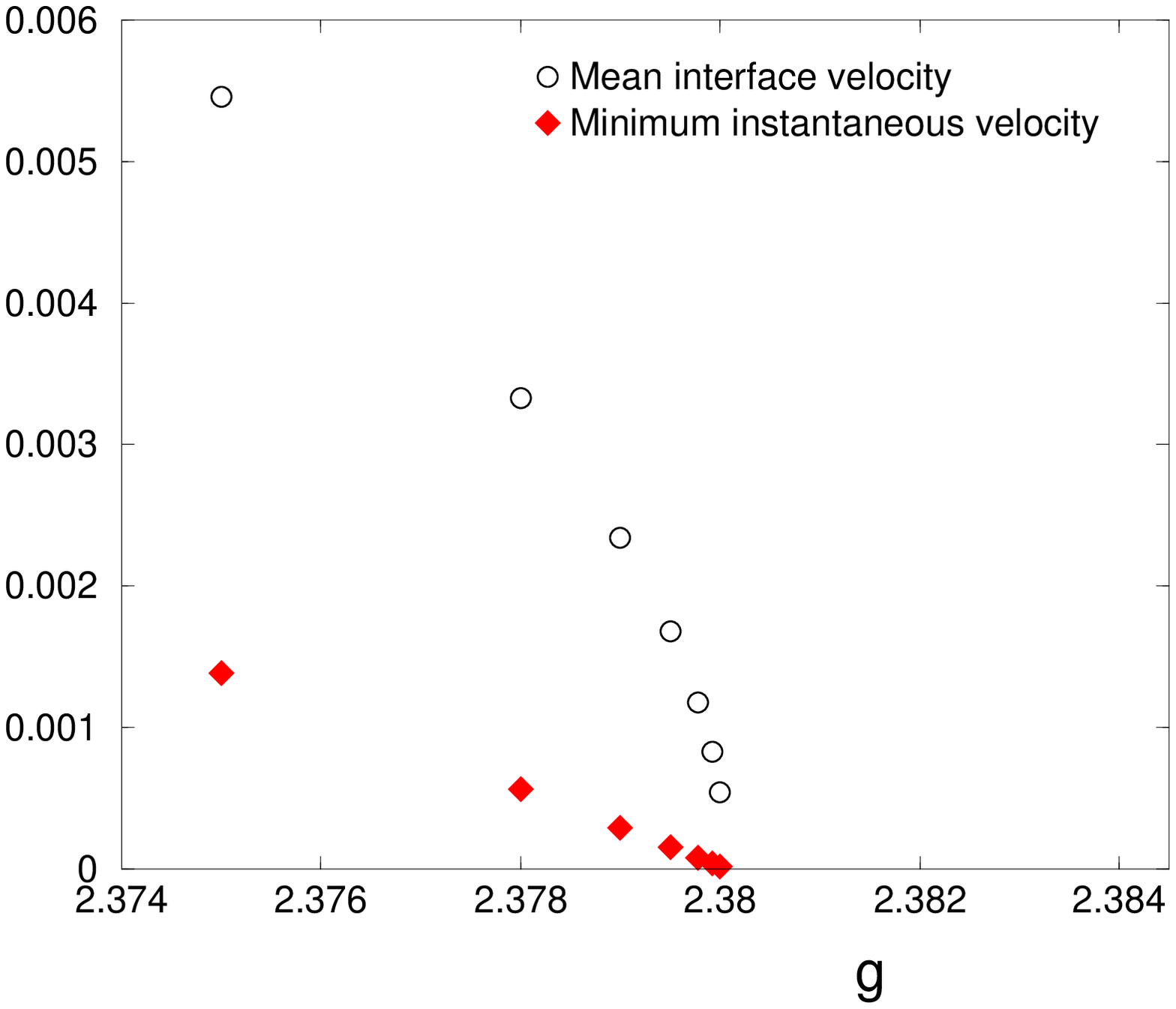,height=6cm}}
 \caption{Mean interface velocity and minimum instantaneous velocity, close to the threshold in
 the moving phase.}
 \label{fig_dstop}
\end{figure}

\subsubsection{A saddle-node bifurcation}

  In figure 5 we have seen that most of the time the interface velocity is close to its
minimum. Its profile differs then little from  the critical one obtained above.

The situation is reminiscent of the one
 encountered in simple models \`{a} la Pomeau-Manneville
 of intermittency  \cite{Pomeau.Intermittency.79}. 
These consider mappings  of the form   
\begin{equation}
 X(t+1) = F_{g}(X(t)),
\label{eqn:intermittent}
\end{equation}
and
describe the laminar-to-intermittent transition  
as a saddle-node bifurcation, i.e., the
merging of two fixed points of (\ref{eqn:intermittent})
into a double fixed point $X^*$, at the critical value of the control parameter. 
In the intermittent phase no fixed point exists, 
the system spends most of its time in a near-critical laminar regime, 
where $X(t) \simeq X^*$, 
with short turbulent bursts during which $X(t)$ varies rapidly.

It is shown in the Appendix that close to threshold 
the present system may indeed be cast in a form 
similar to (\ref{eqn:intermittent}):
 the instantaneous velocity may  be expressed to leading
 orders as a  function of the average position $\bar {z} (t)$ only. 
It can be expanded around a minimum as
\begin{equation}
 v(t) =  v_{min}
       + C \; (\bar {z} (t) -  z_{min})^2 + \ldots ,
\label{eqn:vtmin} 
\end{equation} 
where the minimum velocity
$ v_{min}$  is  given  to leading order by
\begin{equation}
v_{min} = A(g_c -g) + \ldots ,
\label{eqn:zmm}
\end{equation}
 $ z_{min}$ is a position for which the minimum velocity is reached
and the constants $A$ and $C$ can be calculated explicitly (see the Appendix). 

The mean velocity $\bar {v}$ 
may be obtained by integrating (\ref{eqn:vtmin}) over half
the period T, assuming that the region of high velocity makes a
negligible contribution to the total transit time. 
One has 
\begin{equation}
\frac{T}{2} = \int_{z_{min}}^{z_{min} + 1/2} \frac{dz}{v(t)}
 \simeq \int_{z_{min}}^{z_{min} + 1/2} \frac{dz}{v_{min} + C (z - z_{min})^2} \ .
\label{eqn:period} 
\end{equation}
 Hence, for  $v_{min} \ll 1 $ :
\begin{equation}
\bar {v} =  1/ T  \simeq  (C v_{min})^{1/2} / \pi  
\end{equation}
\begin{equation}
\bar {v}     \simeq   \frac{\sqrt{A C}}{\pi} (g_c -g)^{1/2} ,
\label{eqn:vbar}  
\end{equation}
 in agreement with the numerically observed behaviour
and yielding the exponent  of the mean velocity in (\ref{eqn:theta})
\begin{equation}
\theta =  1/2.
\end{equation}

\section {Discussion and conclusion}

 We now discuss the result obtained here  for the
depinning exponent of the infinite-range Leschhorn model
and its relation with other  sytems.
 The value $\theta = 1/2$ is the same as for 
 a contact line on a smooth substrate 
periodically modulated  in the direction of motion
 \cite{Raphael.Dynamics.89, Joanny.Motion.90},
so the picture which emerges from our results is that   very close to threshold
the interface dynamics  reduces to the motion of the center of mass 
in a smooth effective washboard-type periodic potential close
to the critical tilt angle. 

   It differs however from  
the value $\theta = 1$ usually quoted in the literature 
for the depinning exponent at the upper critical dimension
 \cite{Nattermann.Renorm.92, Kardar.dynamics.98},
 and it is natural to ask
 what features of the  model may explain this difference.  

For models with continuous space and continuous relaxational dynamics
 the mean-field behaviour  depends on the analytic
properties  of the pinning potential
 \cite{Narayan.CDW.92, Narayan.Interfaces.93} - 
 for potentials with random cusps (corresponding to discontinuous 
random forces such as assumed in the Leschhorn model) 
the value of the critical exponent is found to be $\theta = 1$.
Close to the upper critical dimension the RG analysis indicates 
that under coarse graining cusp singularities are dynamically generated in the
pinning potential, even if they were absent initially 
\cite{Nattermann.Renorm.92, Chauve.Creep.2000}, 
 so  this case of cusped potentials  
 is  the natural starting point for an $\epsilon$-expansion.

The origin of the difference with the exact result obtained here 
may be traced back to a basic assumption  
 usually made in the mean-field theories 
which  is not fulfilled here
 - namely that in the moving phase the instantaneous velocity $v(t)$
may be replaced
 in the  equations of motion  by its mean value over time,   
which can then be determined
as the solution of a self-consistency equation. 
 This assumption was first introduced   
in the study of the depinning of charge-density  waves 
\cite{Fisher.Density.85}, 
it amounts to saying that there is no qualitative difference between a
situation where the interface is driven at constant velocity and one where 
it is submitted to a constant external force. 
The physical idea  is that
the fluctuations should average out for a very large system, 
and it 
is generally accepted  that non-uniformity 
effects  are irrelevant for the critical mean-field behaviour.

On the contrary, in the model studied here, the instantaneous 
 velocity remains  strongly non-uniform in the thermodynamic limit (see Figure~5),
and the mean and the minimum velocity vanish with different 
exponents. 
As shown by the periodicity of the motion, this is related to 
the discreteness of the allowed positions for the interface, 
through the constraint of unit moves.  
In other words  a periodic modulation of the potential in which 
the interface moves may be a relevant perturbation, in the RG sense.
This is physically reasonable, as above the upper critical dimension
an interface remains flat
 \cite{Nattermann.Renorm.92} 
(in the sense that its mean square width does not diverge with its size), 
so it cannot effectively average out the
underlying  periodic potential. 

 This situation is reminiscent of the pinning of interfaces by the lattice
potential in crystalline materials, first studied by Cahn long ago
\cite{Cahn.Crystal.60}.
The assumption of discrete jumps is relevant to various experimental situations 
where the pinning defects  are well separated  
for instance in dilute magnetic materials or in the motion of contact lines 
on controlled heterogeneous substrates 
\cite{Fermigier.Contact.97, Prevost.Contact.99}, 
 so the properties of that class of models are interesting in their own. 
 The simultaneous effects of disorder and crystal-lattice pinning 
on the static roughening 
of elastic manifolds have been studied by several authors 
(see \cite{Bouchaud.Roughen.92}, \cite{Nattermann.Roughen.99},
\cite{Hazareesing.Rough.98},
\cite{Duxbury.Intermittence.00} and references therein).

 For a different model, 
aimed at describing the motion of visco-elastic interfaces, 
Marchetti et al. \cite{Marchetti.Visco.00} 
have also  observed (numerically) that  in
the  infinite-range limit the velocity fluctuations
do not vanish for large system sizes.
This came as a surprise, which  they attributed 
to an instability of the constant-$v$ solution in the thermodynamic limit.
It  is striking that their system shares this type of
behaviour with our model, 
suggesting that the non-uniformity of the velocity 
may be more general.

\bigskip

We thank B. Carvello, P. Chauve, V. Hakim,  T. Giamarchi,
A. Hazareesing, P. LeDoussal,
 M. M\'{e}zard, T. Nattermann, A. Prevost, E. Rolley and I. Webman 
for stimulating discussions.

\bigskip

\appendix

\section{Appendix: Derivation of  (\ref{eqn:vtmin}) and (\ref{eqn:zmm})}

An expansion of the form  (\ref{eqn:vtmin}) for the instantaneous velocity
$v(t)$ close to the threshold would follow along well-known lines if the evolution
equations could be put under the form of a one-variable relation
\begin{equation}
 \bar {z} (t+1) = F_{g}(\bar {z} (t)),
\label{eqn:zinterm}
\end{equation}
such that the  equation $z =  F_{g}(z)$ 
 admits a double root for $g =g_c$. 
The situation would  be similar to
the  Pomeau-Manneville theory of intermittency
\cite{Pomeau.Intermittency.79}, in which the transition is described as the merging
of two fixed points, a stable and an unstable one, which
then disappear  when the control parameter $g$ is varied.
 We show here that for the uniform disorder distribution studied in Section (3), 
close to the threshold,   the equations of motion  
(\ref{eqn:lambdat}), (\ref{eqn:cutoffbis})   and (\ref{eqn:zt1}) 
can indeed be cast in a form analogous to (\ref{eqn:zinterm}).

 Let us consider the time evolution when the instantaneous velocity is very small,
so that  $ k_{min}(t) $ and the interface width
may be assumed to remain constant during a large number of consecutive time
steps.  
 Eqs (\ref{eqn:lambdat2}) and (\ref{eqn:zlambda})  can then
be combined in a relation giving the present interface average position 
from its values at the 5 previous time steps.  
Using the notations  
 \begin{equation}
 y(t) = y_{0} = \frac{1}{2}(1 - \frac{k_{min} - \bar {z}(t)}{g}), \quad 
 y_{j} = \frac{1}{2}(1 - \frac{k_{min} - \bar {z}(t-j)}{g}), 
 \label{eqn:ymm}
 \end{equation}
for $j= 1$ to $5$,
one gets %
\begin{equation}
g (2 y_{0} -1) =  y_{1} + y_{2}(y_{1}-u) + y_{3}(y_{2}-u)(y_{1}-2u)
 + \cdots + y_{5}(y_{4}-u) \ldots (y_{1}-4u),
\label{eqn:evolution}
 \end{equation}
 where $u = 1/2g$. 
This relation is a five-dimensional dynamical system for $y(t)$, 
it is the time-dependent 
counterpart of  (\ref{eqn:consistency}) and is linear with
respect to each of the $y_{j}$.
It is exact in the pinned phase near the stopping point. 
In the moving phase it is valid locally, close  to a velocity minimum,
and remains valid as long as $k_{min}$ does not change 
and the width of the interface
remains equal to 5.

Since  $v(t)$ and its derivatives are 
assumed to be very small,
we may approximate the interface displacements in (\ref{eqn:evolution}) by
\begin{equation}
y_{j}(t) -y(t) = -\frac{j}{2g}v(t) +\frac{j(j-1)}{4g} \frac{dv}{dt} + \cdots, 
\label{eqn:yj} 
\end{equation}
 where the neglected terms depend on higher-order
derivatives of $v(t)$.  
In the pinned phase $(g > g_c)$, (\ref{eqn:evolution}) admits 
a fixed point where $y_{j} \equiv y_{c}(g)$,
corresponding to the halted interface.
Expanding (\ref{eqn:evolution}) to first order in the
small quantities  $v(t)$ and  $[y_{c}(g) - y(t)]$,
 we obtain 
a linear relation  of the form (\ref{eqn:vslowing}), 
where the proportionality
coefficient vanishes for $ g= g_c$, as
expected from the numerical results in Section~3.

In the moving phase (\ref{eqn:evolution}) has no fixed point 
and we have to expand it around the current value of $y(t)$
(note that as we consider the vicinity of a minimum of $v(t)$, 
this insures that  $dv(t)/d(t) << v(t)$).
Regrouping the terms independent of $v$ and those linear in $v$, 
(\ref{eqn:evolution}) becomes
\begin{equation}
R_{g}(y) \; \simeq \; \frac{v(t)}{2g} \; Q_{g}(y), 
\label{eqn:Qg} 
\end{equation}
 where $R_{g}(y)$ is the polynomial appearing in the study of static solutions 
 and is given by eq.(\ref{eqn:polynome}). 
$Q_{g}(y)$ is a polynomial of degree $4$  in $y$, which may be expressed as 
\begin{eqnarray}
Q_{g}(y) &=& 1+ y(y-u)(\frac{2}{y} + \frac{1}{y-u}) + y(y-u)(y-2u)
(\frac{3}{y} + \frac{2}{y-u} + \frac{1}{y-2u}) \nonumber \\ 
 & & + \cdots + y(y-u) \ldots (y-4u) (\frac{5}{y}+\frac{4}{y-u}+ \ldots  +
\frac{1}{y-4u}). 
\label{eqn:Qy} 
\end{eqnarray}
Using the definition of $y$ (\ref{eqn:ymm}), relation (\ref{eqn:Qg})
 may be cast in the
canonical form (\ref{eqn:zinterm}), justifying the claim made in Section (3). 

Using the fact that $R_{g}(y)$ has a double root for $g= g_c$, i.e., 
\begin{equation}
R_{g_c}(y_c) = \partial R_{g_c}(y)/ \partial y \mid _{y= y_c} =0,
\label{eqn:Rcgc}
\end{equation}
the leading terms in the expansion of $R_{g}(y)$ near the threshold,
in the range where $ y - y_c \; \simeq \; (g-g_c)^{1/2}$, read
\begin{equation}
R_{g}(y) = (\frac{\partial R_{g}}{\partial g})^{\ast} (g- g_c) 
+ \frac{1}{2} (\frac{\partial ^2  R_g}{\partial y^2 })^{\ast} (y- y_c)^2 + \cdots,
\label{eqn:Rg} 
\end{equation}
where for shortness $y_c = y_c(g_c)$ 
and the star symbol denotes quantities evaluated at the threshold.

 Let $\bar {z_c}$ denote the average position  of 
 a halted critical interface such that its trailing edge
 coincides with $k_{min}(t)$.
Then $ y- y_c = (\bar {z} - \bar {z_c})/ 2g $, 
and from (\ref{eqn:Qg}) and (\ref{eqn:Rg})  the leading terms 
in the expansion of $v(t)$ are
of the form
\begin{equation}
 v(t) = A \; (g_c -g) 
       + C \; (\bar {z} (t) - \bar {z_c})^2 + \ldots 
\label{eqn:vtt} 
\end{equation}
The numerical values of the coefficients are
\begin{eqnarray}
A  &=& - \: (\partial R_g/ \partial g)^{\ast} \frac{2g_c}{Q^{\ast}}  \quad
= 0.273786...    \\
C  &=& (\partial ^2  R_g/ \partial y^2)^{\ast} \frac{1}{4 g_c Q^{\ast}} \quad
= 0.167239...
\label{eqn:AC} 
\end{eqnarray}
where  
 Eq.(\ref{eqn:vtt})
 can be cast in the desired form (\ref{eqn:vtmin}),
with the minimum velocity given  by 
\begin{equation}
 v_{min}  =  A(g_c -g) + \ldots   
\label{eqn:zBC}
\end{equation}
as announced in (\ref{eqn:zmm}).
Finally, from (\ref{eqn:vbar}), the critical behaviour of the mean velocity
is
\begin{equation}
\bar {v} \; \simeq \; \frac{\sqrt{A C}}{\pi} (g_c -g)^{1/2} \; \simeq  \; 0.06811
\ldots \; (g_c -g)^{1/2} ,
\label{eqn:vbar2}  
\end{equation}
in good agreement with a numerical study very close to threshold.

\bigskip

\end{document}